\documentclass[twocolumn,english,aps,PRB,groupedaddress,amsmath,floats]{revtex4}
\usepackage[T1]{fontenc}
\usepackage[latin9]{inputenc}
\usepackage{graphicx}



\usepackage{babel}

\begin{document}

\title{Exciton-phonon Bound Complex in Single-walled Carbon Nanotubes Revealed by High-field Magneto-optical Spectroscopy}

\author{Weihang Zhou,$^{1}$ Tatsuya Sasaki,$^{1,4}$ Daisuke Nakamura,$^{1}$ Hiroaki Saito,$^{1,4}$ Huaping Liu,$^{2,3}$ Hiromichi Kataura,$^{2,3}$ and Shojiro Takeyama$^{1}$}

\email{takeyama@issp.u-tokyo.ac.jp}

\affiliation{$^{1}$The Institute for Solid State Physics, The University of Tokyo, 5-1-5, Kashiwanoha, Kashiwa, Chiba 277-8581, Japan}

\affiliation{$^{2}$Nanosystem Research Institute, National Institute of Advanced Industrial Science and Technology, Tsukuba, Ibaraki 305-8562, Japan}

\affiliation{$^3$Japan Science and Technology Agency, CREST, Kawaguchi, Saitama 330-0012, Japan}

\affiliation{$^4$Department of Applied Physics, The University of Tokyo, Hongo 113-8656, Japan}

\date{\today}

\begin{abstract}
High-field magneto-optical spectroscopy was conducted on highly-selected chiral (6,5) specific single-walled carbon nanotubes. Spectra of phonon sidebands in both 1st and 2nd sub-bands were observed to be unchanged by the application of an external magnetic field up to 52 T. Our analyses led to the conclusion that both phonon sidebands in respective sub-band originate from the dark K-momentum singlet (D-K-S) excitons. Moreover, while the relative ordering between the band-edge bright exciton and its zero-momentum anti-bonding counterpart was found to be opposite for the 1st and 2nd sub-bands, the relative ordering between the D-K-S exciton and the band-edge bright exciton was clarified to be the same for both sub-bands. Energy of these D-K-S excitons was estimated to be $\sim$ 21.5 and $\sim$ 37.3 meV above the band-edge bright exciton for the 1st and 2nd sub-bands, respectively.
\end{abstract}


\maketitle

It is nowadays well recognized that phonon mediated transitions play an important role in understanding the optical properties of single-walled carbon nanotubes (SWNTs) \cite{Perebeinos2005, Htoon2005, Plentz2005, Berciaud2010}. The unique one-dimensional sp$^2$-bonded carbon structure of SWNTs leads to exceptionally large excitonic effect as well as enhanced exciton-phonon interaction. Such exciton-phonon bound complexes were predicted theoretically and observed experimentally by means of various spectroscopic techniques, such as photoluminescence, photoluminescence excitation and optical absorption measurements \cite{Perebeinos2005, Htoon2005, Plentz2005, Berciaud2010, Miyauchi2006, Lebedkin2008, Blackburn2012, Zhou2013}. However, it is only recently that these phonon sidebands were confirmed to originate from the dark K-momentum singlet (D-K-S) exciton $|KK^{'}\&K^{'}K>$, among the complicated excitonic structures of SWNTs that comprise a total of 16 excitonic states. \cite{Blackburn2012, Zhou2013}.

Although the efforts so far have unambiguously confirmed the existence of such exciton-phonon bound complex and their mechanism as originating from the D-K-S excitons, it should be pointed out that our understanding on SWNT vibronic sidebands is still far from complete. So far, phonon sidebands of SWNTs have been observed from the lowest sub-band (E$_{11}$) up to the forth sub-band (E$_{44}$) transitions \cite{Berciaud2010}. However, only the E$_{11}$ phonon sideband was confirmed to be associated with the D-K-S exciton. A convincing confirmation for the phonon sideband origin of higher-order sub-band transitions is still missing. Furthermore, our recent magneto-spectroscopic study demonstrates that the relative ordering between the bright $\Gamma$-momentum singlet (B-$\Gamma$-S) exciton $|KK-K^{'}K^{'}(+)>$ and the dark $\Gamma$-momentum singlet (D-$\Gamma$-S) exciton $|KK-K^{'}K^{'}(-)>$ for the (6,5) SWNTs is opposite for the 1st and 2nd sub-bands \cite{Zhou2013R}. This phenomenon is far beyond the expectation of existing theories. More importantly, it leads to the surmise that phonon sidebands of higher-order sub-band transitions may originate from some other excitonic states. On the other hand, it is worth pointing out that, for optical and electronic properties of SWNTs, the D-K-S exciton is as much important as the D-$\Gamma$-S exciton that locates just below the B-$\Gamma$-S exciton. The D-K-S exciton is close in energy with the B-$\Gamma$-S and D-$\Gamma$-S excitons and is always thermally populated at room temperature. Thus, it is interesting for both fundamental physics and device application to investigate the D-K-S excitonic level and its response to external magnetic fields, by means of phonon sideband spectroscopy. To clarify these issues, we perform precise high-field magneto-optical measurements on the E$_{22}$ vibronic sideband of the (6,5) species. The reason to choose the (6,5) species is that, this species has been demonstrated in our previous work to show sub-band dependence for the relative ordering between B-$\Gamma$-S and D-$\Gamma$-S excitons \cite{Zhou2013R}, and their E$_{11}$ phonon sideband has been confirmed to originate from the D-K-S exciton \cite{Blackburn2012, Zhou2013}. It is thus convenient to explore the robustness of the D-K-S excitonic level and the possibility of sub-band dependence for the phonon sideband origin by studying the E$_{22}$ manifold and comparing with the well-established E$_{11}$ results. To the best of our knowledge, this is also the first experimental confirmation for the origin of phonon sidebands other than the E$_{11}$ manifold.

The highly chirality-selected (6,5) SWNTs used in this work were isolated from a high-pressure carbon monoxide (HiPco)-grown mixture by the single-surfactant multicolumn gel chromatography technique \cite{Liu2011}. High degree of tube alignment was realized by embedding SWNTs into polyvinyl alcohol (PVA) film, followed by a stretching process of the SWNTs/PVA film with a stretching ratio of $\sim$ 5. From the optical anisotropy measurements, we extracted the average angle $\theta$ between SWNT axis and the stretched direction to be $\sim 30^{\circ}$. For magneto-absorption measurements, the pulsed magnetic field was generated by a wire-wound solenoid coil with a maximum field of $\sim$ 52 T. Both the incident and transmitted light from a pulsed halogen lamp was guided by optical fibers and finally detected by an electrically cooled ICCD. Spectra were taken on the top of the pulsed field with an exposure time of 5 ms under the Voigt geometry \cite{Zhou2013}.

The main idea we used in this work to investigate the origin of phonon sidebands is based on the distinctive evolutions of relevant excitonic states in magnetic fields. We note that the magnetic field dependent behaviors of phonon sidebands are correlated to their parent excitonic states. Thus, the distinctive evolution of SWNT excitonic states in external magnetic fields, which is well-known as a manifestation of the Aharonov-Bohm (AB) effect \cite{Zaric2006, Ando2004, Ando2006, Takeyama2011, Zaric2004, Ajiki1993, Mortimer2007, Srivastava2008}, serves as an efficient tool to clarify the origin of phonon sidebands. In the following, we would show that both energy and intensity of the E$_{22}$ phonon sideband are immune to the external magnetic field, the same as its E$_{11}$ counterpart \cite{Zhou2013}. However, unlike conventional absorption measurements, the absorption intensity of E$_{22}$ phonon sideband is as low as $\sim$ 0.019 O.D. (optical density) due to dilute dispersion of carbon nanotubes in the stretched PVA film. In a pulsed magnetic field operation, under the restriction of a one-shot measurement with a very short exposure time, it is very difficult to evaluate the degree of changes for both intensity and energy of such a weak absorption peak. Moreover, this weak peak is further obscured by large background signals. To overcome these difficulties, we take the zero-field spectrum every time before we apply the pulsed field and use it as a reference to calibrate the field-dependent spectra. All data used in the analyses were calibrated precisely in this way and confirmed by repeated measurements at each magnetic field.

Typical magneto-absorption spectra for the E$_{22}$ phonon sideband of the stretch-aligned single-chirality (6,5) SWNTs are shown in Fig. 1. Spectra taken at different magnetic fields have been carefully calibrated and slightly offset to have the same baseline level. At zero-field, besides the E$_{22}$ main exciton peak, the phonon sideband was clearly identified to be $\sim$ 215 meV above the main peak. When an external magnetic field was applied, the main exciton peak shows rapid decrease in peak height and obvious peak broadening, as a result of the AB effect. However, it is interesting to note that the phonon sideband at different magnetic fields shows complete overlapping, in strong contrast with the main exciton peak.

For quantitative analyses, it is necessary to extract the evolution of the phonon sideband at each magnetic fields. To do this, an appropriate fitting is needed. However, SWNT absorptive phonon sidebands exhibit unique asymmetric lineshape that results from convolution of the exciton and phonon densities of states \cite{Blackburn2012, Vora2010, Torrens2008}. To account for this unique lineshape, Vora and Torrens \emph{et al}. calculated the lineshape numerically by solving the Bethe-Salpeter equation \cite{Vora2010, Torrens2008}. Here, we utilized the exponentially modified Gaussian (EMG) function \cite{Hanggi1985, Naish1988}, as employed by Blackburn \emph{et al}. \cite{Blackburn2012}, to estimate the intrinsic peak energy of the phonon sideband. Typical result of the fitting using the EMG function is shown in Fig. 2, together with the best-fit curve using the conventional Gaussian function. From the figure, it can be seen clearly that the EMG function gives excellent fitting to the asymmetric lineshape of the E$_{22}$ phonon sideband. From these EMG fittings, we estimated the energy position of the observed phonon sideband in different magnetic fields, as shown by the half-filled circle in Fig. 3(a).
\begin{figure}[t!]
\includegraphics[width=8.5cm]{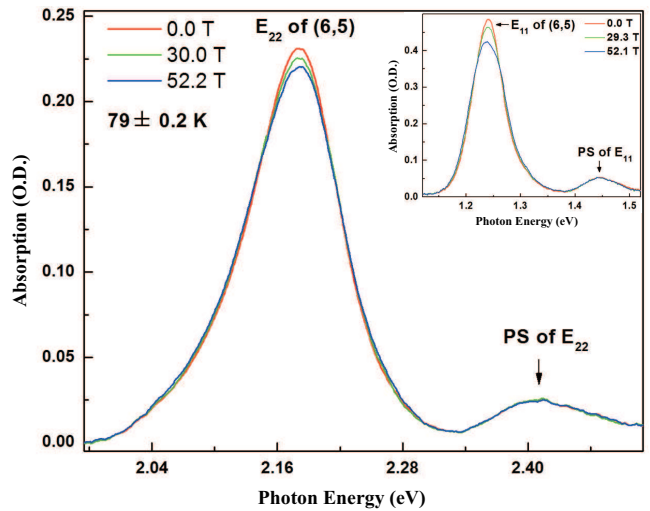}
\caption{(color online) Typical magneto-absorption spectra for the E$_{22}$ phonon sideband of the stretch-aligned (6,5) SWNTs. Light polarization and magnetic field \emph{B} were parallel to the stretched direction of the SWNTs/PVA film. Inset: typical magneto-absorption spectra of the E$_{11}$ phonon sideband \cite{Zhou2013}. PS: phonon sideband.} \label{LDAorbital}
\end{figure}

Theoretical studies show that each sub-band manifold of SWNTs comprises a total of 16 excitonic states \cite{Ando2004, Ando2006}, among which the D-K-S and B-$\Gamma$-S excitons are two most probable candidates for the origin of phonon sidebands \cite{Lebedkin2008, Blackburn2012, Vora2010, Torrens2008, Matsunaga2010, Murakami2009, Kiowski2007}. In the case of the 1st sub-band transitions, there has been intense debate about the origin of the phonon sidebands, and it was confirmed recently that the D-K-S exciton is responsible for the observed phonon sidebands \cite{Blackburn2012, Zhou2013}. In the following, we would follow the procedure as employed in our previous work \cite{Zhou2013} to analyze the origin of the E$_{22}$ phonon sideband.

\begin{figure}[t!]
\includegraphics[width=8.5cm]{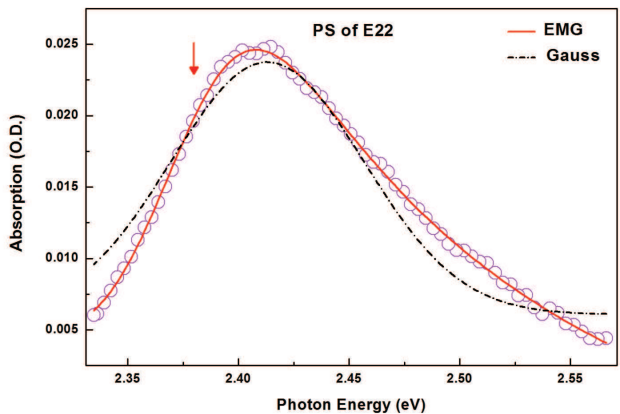}
\caption{(color online) E$_{22}$ phonon sideband absorption spectrum in the absence of magnetic field (open circle) and its best-fit curves using the EMG (solid line) and Gauss (dash-dotted line) functions. The arrow denotes the intrinsic peak position estimated from the EMG fittings.} \label{LDAorbital}
\end{figure}
In the case of the D-K-S exciton, phonon sidebands were believed to originate from the D-K-S exciton coupling with the zone-boundary A$_{1}^{'}$ symmetry phonon \cite{Blackburn2012, Vora2010, Torrens2008, Matsunaga2010, Murakami2009}. Energy of this D-K-S exciton (E$_{\text{K}}$) remains constant in magnetic fields within the available field range \cite{Ando2006}. Moreover, energy of the zone-edge A$_{1}^{'}$ symmetry phonon (E$_{A_{1}^{'}}$) does not change within the available magnetic fields, either. Therefore, energy of the phonon sideband $E_{\text{PS}} = E_{\text{K}} + E_{A_{1}^{'}}$ remains unchanged with and without magnetic field if it originates from the D-K-S exciton. The expected field-dependent energy position of the phonon sideband arising from the D-K-S exciton is shown by the dashed line in Fig. 3(a).

 In the case of the B-$\Gamma$-S exciton, its coupling with the zone-center $\Gamma_{\text{LO}}$ phonon was believed to be the origin of the observed phonon sideband \cite{Lebedkin2008, Kiowski2007}. Evolution of the E$_{22}$ B-$\Gamma$-S exciton in magnetic fields has been identified in our previous work in ultra-high magnetic fields up to 190 T \cite{Zhou2013R}. Its energy evolution can be described well by the empirical formula: $E_{\text{B}\Gamma\text{S}} = E_{\text{g}} - \sqrt{\Delta_{\text{bd}}^{2} + \Delta_{\text{AB}}^{2}}/2$, with $\Delta_{\text{bd}} = - (9.0 \pm 2.5)$ meV, $\mu \approx 0.65$ meV/T ($\Delta_{\text{AB}} = \mu B \cos \theta$ is the so-called AB splitting). Using the reported energy of $\Gamma_{\text{LO}}$ phonon ($E_{\text{LO}} \approx 197$ meV \cite{Blackburn2012}), the energy shift of the phonon sideband $E_{\text{PS}} = E_{\text{B}\Gamma\text{S}} + E_{\text{LO}}$ could be deduced, as shown by the dash-dotted line in Fig. 3(a). Compared with the experimental results estimated from the EMG fitting, it is clearly demonstrated that the observed energy evolution of the phonon sideband agrees quite well with that generated by the D-K-S exciton. This result justifies that the E$_{22}$ phonon sideband also originates from the D-K-S exciton coupling with zone-boundary A$_{1}^{'}$ symmetry phonon.

\begin{figure}[t!]
\includegraphics[width=8.5cm]{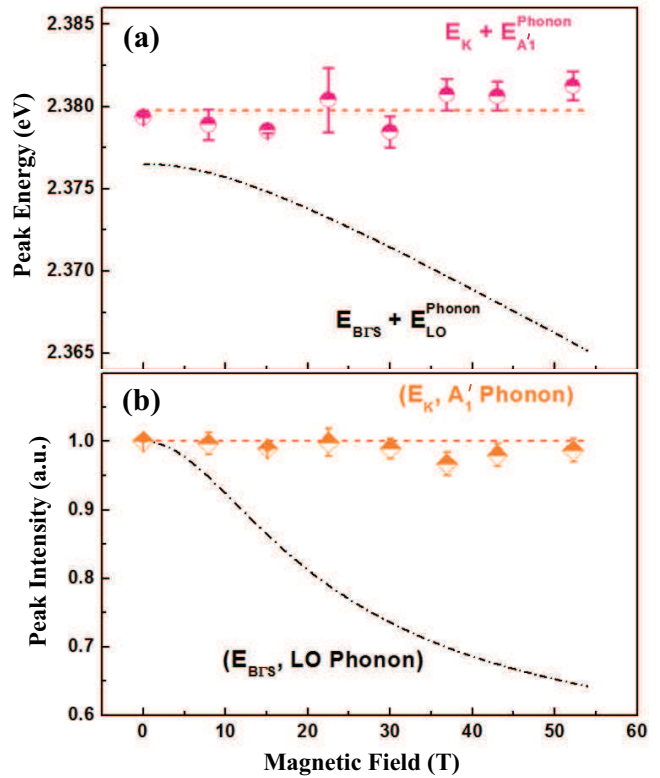}
\caption{(color online) (a) Magnetic field dependent energy of the E$_{22}$ phonon sideband and theoretical predictions. The dashed and dash-dotted lines denote the predicted energy of E$_{22}$ phonon sideband generated by the D-K-S and B-$\Gamma$-S excitons, respectively. (b) Magnetic field dependent intensity of the E$_{22}$ phonon sideband and theoretical predictions. The dashed and dash-dotted lines denote predictions for phonon sidebands generated by the D-K-S and B-$\Gamma$-S excitons, respectively. Intensities have been normalized to their corresponding zero-field intensities, respectively.} \label{LDAorbital}
\end{figure}
The distinctive difference in the magnetic field dependent energies of the D-K-S and B-$\Gamma$-S excitons helped to clarify the origin of the E$_{22}$ phonon sideband. Moreover, the different behavior in the field dependent oscillator strength of the D-K-S and B-$\Gamma$-S excitons as shown in Fig. 3(b) is another confirmation of the sideband origin. As discussed in our previous work \cite{Zhou2013R}, the external magnetic field redistributes the oscillator spectral weight between the B-$\Gamma$-S and D-$\Gamma$-S excitons \cite{Zaric2006, Ando2004, Ando2006, Takeyama2011, Zaric2004, Ajiki1993}. The intensity evolution of the B-$\Gamma$-S exciton can be well described by a simple two-level model: $I_{\text{B}\Gamma\text{S}} = 1/2 + \Delta_{\text{bd}} / 2\sqrt{\Delta_{\text{bd}}^2 + \Delta_{\text{AB}}^2}$ \cite{Zhou2013R}. Considering the fact that an external magnetic field does not change the coupling strength between excitons and phonons within an available range of the field, the intensity evolution of phonon sidebands should follow exactly those of their parent excitons. Using parameters deduced for the B-$\Gamma$-S exciton in our previous work \cite{Zhou2013R}, the intensity evolution of phonon sideband generated by the B-$\Gamma$-S exciton is plotted by the dash-dotted line in Fig. 3(b). Whereas in the case of the D-K-S exciton, the response to external magnetic field is quite different. Theoretical studies show that the D-K-S exciton remains optically inactive with and without magnetic fields \cite{Ando2006}. Based on the fact that changes of coupling strength between excitons and phonons are negligible within available magnetic fields, the intensity of phonon sideband should remain unchanged in external fields if it originates from the D-K-S exciton, as shown by the dashed line in Fig. 3(b). Here, it should be pointed out that, since relevant calculations are mainly focused on the 1st sub-band transitions \cite{Ando2006}, we assume at the beginning that the D-K-S excitons of higher-order sub-bands have the same magnetic field dependent behaviors as their counterpart of the 1st sub-band. Experimental results from EMG analyses are shown by rhombus. By comparison, we can see unambiguously that the intensity evolution follows exactly the prediction of the D-K-S exciton. This confirms undoubtedly, again, that E$_{22}$ phonon sideband originates from the D-K-S exciton. For readers' convenience, typical magneto-absorption spectra of the E$_{11}$ transition were given in the inset of Fig. 1, which clearly shows that phonon sidebands of both sub-bands have the same magnetic field dependent behaviors. The consistency between our independent analyses on the energy and intensity evolutions of the phonon sideband is a strong support for our assumption that the D-K-S excitons of different sub-bands have the same properties.

\begin{figure}[t!]
\includegraphics[width=8.5cm]{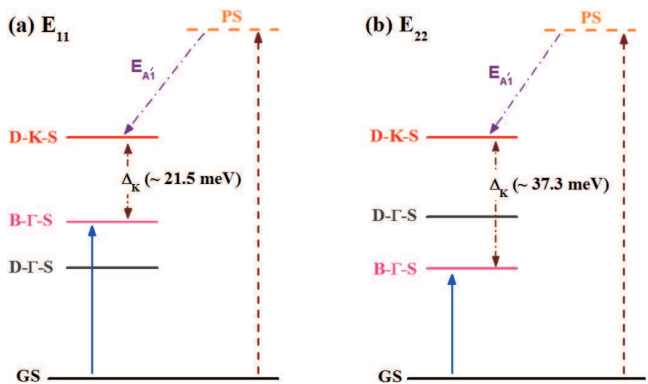}
\caption{(color online) Schematics for the excitonic structure and phonon sideband origin of the (6,5) SWNTs. (a) E$_{11}$ transition, B-$\Gamma$-S exciton lies above the D-$\Gamma$-S exciton. (b) E$_{22}$ transition, B-$\Gamma$-S exciton lies below the D-$\Gamma$-S exciton. Solid arrow: optical transition involving the bright exciton. Dashed arrow: phonon sideband absorption involving the D-K-S exciton. PS: phonon sideband. E$_{A_{1}^{'}}$: energy of the zone-edge A$_{1}^{'}$ symmetry phonon. GS: ground state.} \label{LDAorbital}
\end{figure}
Having confirmed the phonon sideband origin, we can further estimate energy of the E$_{22}$ D-K-S exciton:
\begin{equation}
E_{\text{K}} = E_{\text{PS}} - E_{A_{1}^{'}}
\end{equation}
 With the energy of the A$_{1}^{'}$ symmetry phonon $E_{A_{1}^{'}} = 163$ meV \cite{Blackburn2012}, this calculation returns an energy of the E$_{22}$ D-K-S exciton as $E_{\text{K}} = 2216.8$ meV. Using the known energy of the E$_{22}$ bright exciton, we can further estimate the energy separation of the D-K-S exciton from the B-$\Gamma$-S exciton: $\Delta_{\text{K}}(E_{22}) = E_{\text{K}} - E_{\text{B}\Gamma\text{S}} = 37.3$ meV. Compared with the E$_{11}$ data $\Delta_{\text{K}}(E_{11}) = 21.5$ meV \cite{Zhou2013}, it is interesting to note that the relative ordering between the D-K-S and B-$\Gamma$-S excitons is the same (i.e., positive) for both sub-bands. This is in strong contrast to the case of the D-$\Gamma$-S exciton in the same SWNTs: the relative ordering between the D-$\Gamma$-S and B-$\Gamma$-S excitons is opposite for the 1st and 2nd sub-bands \cite{Zhou2013R}. A schematic diagram summarizing the relative ordering between the D-K-S, D-$\Gamma$-S and B-$\Gamma$-S excitons is shown in Fig. 4, together with the phonon sideband origin for both sub-bands.

In summary, we report high-field magneto-optical study on the exciton-phonon sidebands of the (6,5) species for both 1st and 2nd sub-bands. Our independent analyses on the energy and intensity evolutions come to the same conclusion, that the E$_{22}$ phonon sideband of the (6,5) species originates from the D-K-S exciton, the same as its counterpart of the E$_{11}$ transition. By means of the phonon sideband spectroscopy, we further estimated the relative energy position of the D-K-S, D-$\Gamma$-S and B-$\Gamma$-S excitons. While the relative ordering between D-$\Gamma$-S and B-$\Gamma$-S excitons was found to be opposite for the 1st and 2nd sub-bands, the relative ordering between the D-K-S and B-$\Gamma$-S excitons was clarified to be the same for both sub-bands. Excitonic effect and phonon-mediated transitions are crucial for SWNT optical properties, thus, the clarification of phonon sideband origin and robustness of the D-K-S excitonic state in this work would be helpful to guide future experiments as well as commercial applications.

\acknowledgements{We are obliged to Prof. K. Kindo for supplying a nondestructive pulsed magnet. One of the authors (W. H. ZHOU) thanks for the financial support of the post-doctoral research fellowship at the Institute for Solid State Physics.}

\end{document}